\documentclass{ifacconf} 

\usepackage{amsmath,amssymb,amsfonts}
\usepackage{algorithm}
\usepackage{graphicx,import}
\usepackage{natbib}
\usepackage{textcomp}
\usepackage{xcolor}
\usepackage{bm,nicefrac}
\usepackage{mathrsfs}
\usepackage{algpseudocode}
\usepackage{multirow}
\usepackage{url}
\def\BibTeX{{\rm B\kern-.05em{\sc i\kern-.025em b}\kern-.08em
    T\kern-.1667em\lower.7ex\hbox{E}\kern-.125emX}}
\usepackage{array}
\newcolumntype{L}[1]{>{\raggedright\let\newline\\\arraybackslash\hspace{0pt}}m{#1}}
\newcolumntype{C}[1]{>{\centering\let\newline\\\arraybackslash\hspace{0pt}}m{#1}}
\newcolumntype{F}[1]{>{\let\newline\\\arraybackslash\hspace{0pt}}m{#1}}
\definecolor{tudelft-fuchsia}{cmyk}{0.19,1,0,0.19}
\definecolor{tudelft-cyan}{cmyk}{1,0,0,0}

\newcommand{\RMGF}[1]{#1}
\newcommand{\R}{\mathbb{R}}

\begin{document}

% \maketitle
% \thispagestyle{empty}
% \pagestyle{empty}

\begin{frontmatter}
		
		\title{Beta Residuals: Improving Fault-Tolerant Control for Sensory Faults via Bayesian Inference and Precision Learning} %\thanksref{footnoteinfo}}
		
% 		\thanks[footnoteinfo]{
% 			This work has been partially supported by ...
% 		}
		
		\author[Oxford]{Mohamed Baioumy} 
        \author[Amazon]{William Hartemink}
        \author[Delft]{Riccardo M.G. Ferrari}
        \author[Oxford]{Nick Hawes}
        
        \address[Oxford]{Oxford Robotics Institute, University of Oxford, United Kingdom.
        For correspondence: mohamed@robots.ox.ac.uk.}
        \address[Amazon]{Amazon Web Services, Texas, United States.}
        \address[Delft]{Delft Center for Systems and Control, Delft University of Technology, Mekelweg 2, 2628 CD, Delft, The Netherlands. \\
		(e-mail: r.ferrari@tudelft.nl)}

% 		\author{Alexander J. Gallo$^\diamond$, and }\author{Riccardo M. G. Ferrari$^{\diamond}$}\\
% 		\vspace{.1 in}{\small
% 			{$^\diamond$\it Delft Center for Systems and Control, Mechanical, Maritime, and Materials Engineering, TU Delft, Delft, Netherlands}\\
% 			(email: {a.j.gallo,r.ferrari}@tudelft.nl)
% 		}
		
		\begin{abstract}
%
% This work presents a fault-tolerant control scheme for robotic systems based on active inference. The proposed solution makes use of the sensor prediction errors in the free-energy to simplify the residuals and thresholds generation for fault detection and isolation and does not require additional controllers for fault recovery. To achieve so, we propose a modified version of the active inference controller referred to as `decoupled Active Inference Control'. This improves state-estimation with respect to the classical active inference formulation and it allows to make use of fault detection techniques with probabilistic robustness, reducing the chances of false positives. 
%
Model-based fault-tolerant control (FTC) often consists of two distinct steps: fault detection \& isolation (FDI), and fault accommodation. In this work we investigate posing fault-tolerant control as a single Bayesian inference problem. Previous work showed that precision learning allows for stochastic FTC without an explicit fault detection step. While this leads to implicit fault recovery, information on sensor faults is not provided, which may be essential for triggering other impact-mitigation actions. In this paper, we introduce a precision-learning based Bayesian FTC approach and a novel \emph{beta residual} for fault detection. Simulation results are presented, supporting the use of beta residual against competing approaches. 
\end{abstract}

% \begin{keywords}
% Bayesian Inference, Fault-tolerant control, Robotics.
% \end{keywords}

%		\begin{keyword}
%			Secure control, something else, a third thing. 
%		\end{keyword}
		
	\end{frontmatter}

\section{Introduction}

%%% NEW VERSION

% Background and motivation

The emergence of autonomous robotic systems calls for the introduction of fault-tolerant control architectures that can guarantee safe operation even under faulty conditions.
\emph{Fault-Tolerant Control} (FTC) techniques can limit the impact of faults at process, sensor or actuator level \citep{blanke2000fault,ChengBook}. FTC approaches typically consists of two steps: fault detection, isolation and identification (FDI); and fault accommodation.

% The problem
Model-based approaches to FTC proved to be very powerful \citep{gao2015survey}, but requires high amounts of knowledge and data for modelling the system and designing FDI residuals and thresholds. This holds especially for cases modelled as stochastic systems, where probabilistic rather than deterministic solutions are preferred. %\cite{fang2015stochastic-ft,rostampour2020privatized}}.

% What current literature did to solve it
A notable contribution to solving this problem has been provided by the prolific literature on Bayesian approaches to FDI. \cite{Cai2017-sb} provides an overview of early works which used static Bayesian Networks to model the causal relation between faults and symptoms, and obtain a diagnosis via inference. % \cite{Szolovits1978-oj}.
Later works introduced Dynamic Bayesian Networks (DBN) as a way to encode transient and time varying behaviours, and general spatial and time relations. For instance \cite{Lerner2000-cq} and \cite{Verma2004-vv} use, respectively, a Kalman and a Particle Filter to track possible fault hypotheses, while \cite{Codetta-Raiteri2015-sk} illustrate the use of Probabilistic Graphic Models (PGM) and DBNs for spacecraft FDI. In \cite{Sun2020-ht} a Bayesian Recurrent Neural Network is used as a probabilistic model, whose posterior is approximated via a variational approach.

% What is still unsolved or not satisfying
\RMGF{Still, several challenges remain at the forefront of research on this topic. %\cite{baioumy2020active,pezzato2020active}.
On one side, an integrated Bayesian approach for both FDI and fault recovery is sought. On another, the computational complexity of inference over DBN can be unsuited to embedded implementations on board autonomous robots, when complex models are employed. Finally, Bayesian approaches need adaptation capabilities to cope with time varying systems, for instance due to ageing or other non-fault phenomena.}

\RMGF{In order to address the first two issues, several works in the field of robotics and control have recently taken inspiration from the \emph{active inference} framework. In neuroscience, it is regarded as a general theory for perception and action \citep{friston2} and can be used to understand decision making of biological agents and to build artificial agents. Active inference methods for control have shown promising results in deterministic FTC \citep{Corrado2020iwai, baioumy2021ECC}, stochastic FTC \citep{baioumyIWAI2021}, adaptive control for robot manipulators \citep{buckley,Pezzato2020, baioumy2020active} and state-estimation \citep{lanillos1, friston5}.}
%In particular, \emph{Active Inference Control} (AIC) allows for combined FDI and fault recovery as a single Bayesian inference procedure, and does not require the knowledge of a model of the system's dynamics.}

% What we did to solve it

\RMGF{In this work we address the third challenge as well, by presenting a Bayesian controller for stochastic fault-tolerant control that can track naturally occurring variations and faults via \emph{precision learning}. . In particular, we model the precision (inverse covariance) of each sensor as a random variable and compute on-line its posterior. The \textit{expected precision} acts as a measure of the probability of a sensor being faulty, and its inverse is used to weight a given sensor measurement in the control law. This approach does not require a-priori information about fault characteristics, or an explicitly-defined fault detection criteria and recovery mechanisms.
Still, operators may need an interpretable estimate of the location and probability of a fault. To address this, we introduce a so-called \emph{beta residual} based on the estimation of hyper-parameters of the precision distribution density. A corresponding residual evaluation based on the logistic function is then proposed.}

\RMGF{The main contributions of this paper can then be summarised as
\begin{enumerate}
    \item a set of controllers performing \emph{sensor} FTC as Bayesian inference by means of precision learning;
    \item a novel approach for extracting an \emph{interpretable} quantity, the beta-residual, representing the controller's Bayesian belief of sensor faultiness.
\end{enumerate}
%
% For the sake of simplicity we limit the present analysis to sensor fault accommodation, but the proposed approach can be extended to process and actuator faults.
}

% Paper structure
\RMGF{The rest of the paper is organized as following: Section~\ref{sec:control_models} introduces two types of Bayesian control methods which are used in Section~\ref{sec:FTC_precision_learning} to implement FTC via precision learning. The interpretable beta residual is defined in Section~\ref{sec:beta_residual}, while simulation results and conclusions are presented, respectively, in Section~\ref{sec:results} and \ref{sec:conclusions}.}

\section{Control models}\label{sec:control_models}

In this section, we recall results from prior work to motivate the work proposed in this paper. We first formalize the system dynamics in a problem statement; then, we present two controllers based on Bayesian inference: the unbiased active inference controller (u-AIC) of \citep{baioumy2021ECC, baioumyIWAI2021} and a general Bayesian Controller (BC).

% . These control models will be extended to be fault-tolerant in later sections. 

% The first controller is referred to as the unbiased active inference controller (u-AIC) and was introduced in \cite{baioumy2021ECC}. This is a feedback controller: control actions are generated by comparing observations to the desired state. The controller is unaware of the system dynamics, similar to a PID controller. 

% The second controller is referred to as the Bayesian Controller (BC). This controller is equipped with a dynamics model and can predict future trajectories. It then selects control actions that maximize the probability of future trajectories reaching the (time-varying) goal states.

\subsection{Problem statement}

Let us consider a discrete-time nonlinear system as 
\begin{equation}
    \begin{cases}
        {\bm x}_{t+1} &= f({\bm x}_{t}, {\bm u}_{t}) + \bm q \\
        \bm y_t &= g(\bm x_t) + \bm \eta
    \end{cases} \, ,
\label{eq:discrete_system_dynamics}
\end{equation}
\noindent where $\bm x_t \in \R^{N_x}$, $\bm u_t \in \R^{N_u}$ and $\bm y_t \in \R^{N_y}$ are, respectively, the system's state, input and output vector. The variables $\bm q \in \R^{N_x}$ and $\bm \eta \in \R^{N_y}$ are zero-mean, Gaussian process and measurement noises, instead.
%
% The state is hidden and partial information is provided as sensor measurements $\bm y_t^{(i)}$, with $i \in \{1, \dots , N_y\}$ . We refer to the sensor measurement from all $N_y$ sensors collected at the same time-step as simply $\bm y_t$. Each sensor is affected by zero mean Gaussian noise $\bm \eta^{(i)}$, which gives the observation equation: $$\bm y_t = g(\bm x_t) + \bm \eta.$$
%and visual sensors are affected by barrel distortion \cite{lipkin, cameraNoise}.
The controller aims to steer the system to a goal state $\bm{x}_{\text{goal}}$ by applying a control action $\bm u_t$. At any time, unknown to the controller, sensor faults can occur. This includes \emph{hard faults} (complete failure) and \emph{soft faults} (e.g. arbitrary sensor drifts or constant offsets). 

% \begin{equation}
%     \bm y_t = g(\bm x_t) + \bm \eta.
% \end{equation}

% Consider cruise control in particular 
% $x_{t+1} = f(x_t, u_t) + noise$
% $y_i = g_i(x_t) +  \eta_i$

% Additionally, we discuss two methods 

\subsection{The Unbiased Active Inference Controller}
\label{subsec:unbiased_active_inference_controller}
The (u-AIC) 
% This is a feedback controller, where observations are compared to the reference and the resulting error is used to compute control actions. In particular, placing specific constraints on the controller parameters can make the controller behave similarly to a PID controller. 
% make this paragraph feel more like a summary of what follows
%
uses a generative probabilistic model with the current state $\bm{x}$, control action $\bm{u}$ and observations $\bm{y}$. The system can have multiple observation modalities. For instance, a robot can have joint encoders for position and velocity in addition to a camera. As a running example throughout the paper, we derive the equations for a system with two joint encoders $\bm y^{(1)}, \bm y^{(2)}$ and a visual sensor $ \bm y^{(v)}$. In this case the joint distribution $p(\bm{x}, \bm{u}, \bm y^{(v)}, \bm y^{(1)}, \bm y^{(2)})$ is assumed to be factorized as: 
\begin{equation}
    p(\bm{x}, \bm{u}, \bm y^{(v)}, \bm y^{(1)}, \bm y^{(2)}) = \underbrace{p(\bm{u}|\bm{x})}_{control} \underbrace{p(\bm y^{(v)},\bm y^{(1)},\bm y^{(2)}|\bm{x})}_{observation \hspace{1 mm}model} \underbrace{p(\bm{x})}_{prior} 
    \label{eq: factorization_d-aic_joint_probabilistic_distribution_ with_explicit_actions}
\end{equation}

Given the sensor data, we then aim to find the posterior over states and actions $p(\bm{x}, \bm{u}| \bm y^{(v)}, \bm y^{(1)}, \bm y^{(2)})$. As is common in variational Bayesian inference, we approximate the posterior with a variational distribution $q(\bm{x}, \bm{u})$ and utilize the mean-field assumption ($q(\bm{x}, \bm{u}) = q(\bm{x})q(\bm{u})$) and the Laplace approximation \cite{fox2012tutorial}. The posterior over the state $\bm{x}$ is assumed Gaussian with mean $\bm{{\mu}}_{x}$. The posterior over actions $\bm{u}$ is also assumed Gaussian with mean $\bm{\mu}_{u}$. This results in the expression for \emph{variational free-energy}, a quantity which is then minimized to generate control actions:
\begin{equation}
     F = -\ln p(\bm{\mu}_{u}, \bm{{\mu}}_{x}, \bm y^{(v)}, \bm y^{(1)}, \bm y^{(2)}) + C.
\end{equation}

Detailed derivations of the above are available in \cite{baioumy2021ECC}, while more general information about variational inference is available in \cite{murphy2012machine}. It then follows, assuming Gaussian distributions, that $F$ can be factorized as in eq.~\eqref{eq: factorization_d-aic_joint_probabilistic_distribution_ with_explicit_actions} and then expanded to
\begin{equation}
\begin{split}
    \label{eq:laplace_F_final_vector}
    F  
    &=  \frac{1}{2}(\bm{\varepsilon}_{y^{(1)}}^\top P_{y^{(1)}}\bm{\varepsilon}_{y^{(1)}}
    +  \bm{\varepsilon}_{y^{(2)}}^\top P_{y^{(2)}}\bm{\varepsilon}_{y^{(2)}}
    + \bm{\varepsilon}_{y^{(v)}}^\top P_{y_v}\bm{\varepsilon}_{y^{(v)}}\\
    &+ \bm{\varepsilon}_{x}^\top P_{x}\bm{\varepsilon}_{x}
    + \bm{\varepsilon}_{u}^\top P_{u}\bm{\varepsilon}_{u} - \ln|P_{u}P_{y^{(1)}}P_{y^{(2)}}P_{y^{(v)}}P_{x}|)
    + C,
\end{split}
\end{equation}

\noindent where $\bm{\varepsilon}_{y^{(1)}}$, $\bm{\varepsilon}_{y^{(2)}}$, $\bm{\varepsilon}_{y^{(v)}}$ are the sensor prediction errors of position encoder, velocity encoder, and the visual sensor respectively. Furthermore, $\bm{\varepsilon}_{x}$ and $\bm{\varepsilon}_{u}$ are the prediction errors for the prior on the state and on the control action. 

The prediction error on the state prior $ \bm{\varepsilon}_{x} = (\bm{\mu}_x - \hat{\bm x} )$ is computed from $\hat{\bm x}$, the prediction of the state, which is a deterministic value. The prediction can be computed via the prediction step of a Kalman filter: an advantage of the (u-AIC) is that an accurate model is not required. In this paper a simple random walk is assumed.

% While in such cases the prediction computation would require the knowledge of an accurate dynamic model, in this u-AIC controller propagates forward in time the current state belief by computing the position as the discrete time integral of the velocity, using a first-order Euler approximation. 

%using the simplified discrete time model:

% \begin{equation}
%     \label{eq:euler_integration_a}
%     \hat {\bm x}_{k+1} = 
%     %\begin{bmatrix}
%     %\dot{q}_{k+1}\\
%     %q_{k+1}
%     %\end{bmatrix}
%     %=
%     \begin{bmatrix}
%     I & I \Delta t \\
%     0 & I
%     \end{bmatrix}
%     \bm \mu_{x,k}
%     %\begin{bmatrix}
%     %\dot{q}_{k}\\
%     %q_{k}
%     %\end{bmatrix} \, ,
% \end{equation}
% %
% where $I$ represents an unitary matrix of suitable size.
% This form assumes that the velocities will remain roughly constant and the position is thus computed as the discrete time integral of the velocity, using a first-order Euler approximation.

Finally, the information about the target/goal state is encoded in the control bias distribution $p(\bm u | \bm x)$. We choose this distribution to be Gaussian as well with a mean of $f^*(\bm{\mu}_x, \bm{\mu}_d)$, which is a function that steers the systems toward the target. This then results in  $\bm{\varepsilon}_{u} = (\bm{\mu}_u - f^*(\bm{\mu}_x, \bm{\mu}_d))$. The function $f^*(\bm{\mu}_x, \bm{\mu}_d)$ can be \textit{any} controller, for instance a proportional (P) controller: $f^*(\bm{\mu}_x, \bm{\mu}_d)) = P (\bm{\mu}_d - \bm{\mu}_x)$.

\subsection{Bayesian Controller}

% Both the standard AIC and the unbiased AIC do not require a detailed dynamic model. This has many advantages in adaptive, robust and fault-tolerant control. However, these approached can not plan since they are not equipped with a model. This is discussed in \cite{2020ICRA_baioumy}, which presents the foundation for performing model-predictive control. The factor graph is illustrated in \ref{fig:factor_graph_mpc} .

% \begin{figure}[!htb]
%     \centering
%     \includegraphics[width=0.35\textwidth]{img/fg_mpc.JPG}
%     \caption{Model-predictive control structure represented as a factor graph.}
%     \label{fig:factor_graph_mpc}
% \end{figure}

% The model for the predictive controller includes the current state $\mathbf{s_{t}}$, the future state $\mathbf{s_{t+1}}$, the control action $a_t$ and the observation $o_t$. The aim is to compute $p(\mathbf{s_t}, \mathbf{s_{t+1}}, \mathbf{a_t} |\mathbf{o_t})$. Similarly to the reactive case we approximate this distribution with Gaussian variational distributions. The posteriors over states $\mathbf{s_{t}}$ and $\mathbf{s_{t+1}}$ would have the means, $\pmb{{\mu}}_{s}$ and $\pmb{{\mu}}_{s+1}$ respectively. The distribution over actions is also assumed Gaussian with a mean of $\pmb{\mu}_{a}$.

We now consider a Bayesian controller as in 
%\cite{watson2021advancing, levine2018reinforcement, 2020ICRA_baioumy},
\cite{2020ICRA_baioumy}, whose generative model is defined as
% with analytical redundancy. We define a generative model that includes states, control actions and observations. We then put a prior on future states that encodes our target. This is similar to many approaches in frameworks performing control as inference \cite{watson2021advancing, levine2018reinforcement, 2020ICRA_baioumy}. 

\begin{equation}
\begin{split}
\label{eq:Baysian_MPC}
    &p(\bm x_t, \bm x_{t+1}, \bm y_t, \bm u_t) \propto \\ &\underbrace{ p(\bm x_t)}_{\text{Prediction prior}}  
\underbrace{p(\bm x_{t+1}| \bm x_{t}, \bm u_{t})}_{\text{Transition model}} 
\underbrace{p(\bm y_t | \bm x_t)}_{\text{Observation model}} 
\underbrace{ p(\bm x_{t+1})}_{\text{Goal prior}} \,.
% \underbrace{ p(\bm u_{t}).}_{\text{Control prior}} 
\end{split}
\end{equation}
The prediction prior $p(\bm x_t)$ is defined the same way as in the u-AIC. The transition model $p(\bm x_{t+1}| \bm x_{t}, \bm u_{t})$ describes how the agent transitions from a state $ \bm x_{t}$ to a future state $ \bm x_{t+1}$ by taking an action $ \bm u_t$. The observation model $p(\bm y_t | \bm x_t)$ for a given sensor, defines the (noisy) relationship between the states and observation. 

The general approach for this paper works for any choice of distributions. However, for the sake of simplicity, we choose a Gaussian transition model $p(\bm x_{t+1}| \bm x_{t}, \bm u_{t}) = \mathcal{N}(\bm x_{t+1}; f(\bm x_{t}, \bm u_{t}), \Sigma_f)$, where $f(\cdot)$ is a transition function and $\Sigma_f$ is the uncertainty over the transition. This is equivalent to assuming $q \sim \mathcal{N}(\bm 0, \Sigma_f)$ in eq.~\eqref{eq:discrete_system_dynamics}. Similarly, the observation model is chosen to be Gaussian as $p(\bm y_t | \bm x_t) = \mathcal{N}(\bm y_t; g(\bm x_t), \Sigma_y)$.

The unusual element in this model is that the goal state is specified by adding a goal prior over a future state $ \bm x_{t+1}$, which we implicitly predict to reach. The optimal control action leading to this can thus be computed by Bayesian inference.
% To satisfy this condition the agent would have to converge to a control action $\bm u_t$ such that it expects to reach the goal state. The  resulting inferred control action is therefore the optimal control action in this Bayesian formulation.
The goal prior is also chosen to be Gaussian $\mathcal{N}(\bm x_{t+1}; \bm x_{\text{goal}}, \Sigma_{\text{goal}})$. The smaller the value for $\Sigma_{\text{goal}}$ is, the more aggressively the controller will act to reach the target.

% Finally, we could also include a control prior $p(\bm u_{t})$. If that is chosen to be Gaussian as well, this results in 

Solving this model requires computing the target posterior $p(\bm x_t, \bm u_t | y_t)$. Computing $p(\bm x_t | y_t)$ results in filtering and computing $p(\bm u_t | y_t)$ results in control. Since all our distributions are Gaussian, the negative log-likelihood of this model will consist of least-square terms and logarithms. For our model this would lead to four terms since we have four Gaussian factors in eq.~\eqref{eq:Baysian_MPC}. To simplify notation, we define $||x_1 - x_2||^{2}_{\Sigma} = (x_1 - x_2)^\top\Sigma^{-1}(x_1 - x_2) + \ln \Sigma^{-1}$. The full negative log-likelihood for the controller is then:

\begin{equation}
\begin{split}
    &- \mathcal{L}  =  ||\bm x_{t+1} - f(\bm x_t, u_t)||^{2}_{\Sigma_f} + ||\bm y_t - g(\bm x_t)||^{2}_{\Sigma_g} + \\
    &||\bm x_{t+1} - \bm x_{goal}||^{2}_{\Sigma_{g}} +
    ||\bm x_t - \bm \hat{x}||^{2}_{\Sigma_1} = F. 
\end{split}
\label{eq:least_square_loss}
\end{equation}

We can estimate the state $x_t$ and control action $u_t$ by minimizing the negative log-likelihood by using any suitable optimization method.
It follows that the result will depend on the prediction prior ($\bm \hat{x}$) and on the observed value from the sensors $\bm y_t$. Balancing these two terms is equivalent to the prediction and measurement update steps of a Kalman filter, as proved in \cite{ho1964bayesian}. The future state $x_{t+1}$ is constrained by the goal state $x_{\text{goal}}$ (third term in $- \mathcal{L}$). Thus, to minimize the whole expression, the value for the control input $u_t$  needs to make the system evolve, according to the model $f(\bm x_t, u_t)$, towards the goal state.
%This follows from the first term in the expression.

% \begin{rem}
% The key idea of this controller is that the current state is constrained by the prediction step and the measurement, the future state is constrained by the goal prior, and thus the control action will have to move the system towards the latter. 
% \end{rem}

% The belief over $x_t$ and the control action $u_t$ are obtained by any optimization method on the equation \ref{eq:least_square_loss} such as performing gradient descent. Since a dynamical system requires a real value for the control action, the mean of the distribution over actions is used as input (which is the expected value). 

% One might also add a control prior $p(\bm u_t)$ to eq.~\ref{eq:Baysian_MPC} to bias the control action, to introduce an actuator penalty, for example. This will add another quadratic term to $- \mathcal{L}$, as every Gaussian added to the generative model reduces to an additional quadratic cost. Minimizing $- \mathcal{L}$ is similar to minimizing a sum of quadratic costs, as is done for instance in classic LGQ control or model-predictive control. 
\section{Fault-tolerant control using precision learning}\label{sec:FTC_precision_learning}

This section introduces the first major contribution of this paper: achieving FTC with precision learning. 
%Previously, we introduced two controllers: the u-AIC and Bayesian controller. In this section, we show how to expand these controllers by performing precision learning using one of
Two techniquess will be presented: point-based precision learning, as in \citep{baioumyIWAI2021}; and a Bayesian method, which is another novel contribution of this paper.

\subsection{Point-mass approaches}
The observation matrix $\Sigma_{y}$ can be updated via gradient descent on $\mathcal{F}$ as: $\dot{\Sigma}_{y} = -\kappa_{\sigma} \frac{\partial \mathcal{F}}{\partial \Sigma_{y}}$. Such update rule has several practical issues. First, in the present case, it would lead to inverting the possibly high dimensional matrix $\Sigma_{y}$. A work around is to directly update its inverse, the \emph{precision} matrix:

\begin{equation}
    \label{eq:observaion_variance_update}
    \dot{\Sigma}_{y}^{-1} = -\kappa_{\sigma} \frac{\partial \mathcal{F}}{\partial \Sigma_{y}^{-1}}.
\end{equation}

The second issue is that a $\Sigma_{y}$ needs to be positive semi-definite.
% However, the update rules from equations \eqref{eq:observaion_variance_update} may violate these conditions.
A solution is to perform a re-parameterization with a strictly positive function such as an exponential.
% .  i.e. every diagonal element of $\Sigma^{-1}$ is a scalar $\omega$ and we assume that $\omega = \exp{\zeta}$ and we perform gradient descent on $\zeta$: 
%
% \begin{equation}
%     \label{eq:exp_variance_update}
%     \dot{\zeta} = -\kappa_{\zeta} \frac{\partial \mathcal{F}}{\partial \zeta}
% \end{equation}
%
% \noindent where $\kappa_{\zeta}$ is the gradient step-size.

An alternative method is to set a lower bound on the variance, as done in \cite{bogacz2017tutorial}.
% Both methods ensure the variance is positive (positive semi-definite in the case of matrices).

\subsection{Bayesian approaches for one-dimensional problems}
Once again consider a one-dimensional problem where observations $y$ are affected by noise generated by a Gaussian with a known mean $C$ and scalar precision $\omega$, $\mathcal{N}(y; C, \omega)$. Given some observation $y$, we wish to find the posterior $p(\omega|y)$. To do this, we choose the prior over the precision $p(\omega)$ to be a \textit{gamma distribution} defined as: 

$$\Gamma(\omega ; a, b)  = \frac{b^a}{\Gamma(a)} \omega^{a-1}e^{-\omega b}$$

\noindent where $a$ and $b$ are the parameters of the distribution and $\Gamma(a) = (a - 1)!$  is a factorial function. For example, $\Gamma(5) = 4! = 24$. Now to compute the posterior, we multiply the prior with the Gaussian likelihood model of $p(y|\omega)$ and obtain the posterior which is also a gamma distribution as shown below.

$$p(\omega) = \Gamma(\omega ; a, b) \propto \omega^{a-1}e^{-\omega b}$$
$$p(\omega|y) \propto p(y|\omega)p(\omega) \propto \omega^{0.5+a-1}e^{-\omega (b + \frac{(y - C)^2}{2})}$$
$$p(\omega|y) = \Gamma(\omega ; a + \frac{1}{2}, b + \frac{(y-C)^2}{2})$$

The last equation shows a simple update rule to modify the belief over the precision for every observation. In the optimization for the state, the following quantities are used: expected precision $\mathbb{E}[\omega] = a/b$, $\text{Mode}[\omega] = (a-1)/b$ and $\text{Var}[\omega] = a/b^2$. Note that while a gamma distribution was assumed for the precision, as is common in Bayesian inference, this analysis may extend to other distributions.

\subsection{Bayesian approaches for n-dimensional problems}
Using the gamma distribution is limited to one-dimensional problems. For n-dimensional problems, the Wishart distribution is used. Wishart distribution is parameterized by $\mathbf{x}$: a squared positive definite matrix of size $p$ containing random variables, and $\mathbf{V}$ a symmetric positive define matrix of size $p$ as well. Then if $n \geq k$ the Wishart distribution over $\mathbf{x}$ is given by:

$$
p(\mathbf{x})=\frac{1}{2^{n k / 2}|\mathbf{V}|^{n / 2} \Gamma_{k}\left(\frac{n}{2}\right)}|\mathbf{x}|^{(n-k-1) / 2} e^{-(1 / 2) tr \left(\mathbf{V}^{-1} \mathbf{x}\right)}
$$

\noindent where $|\mathbf{x}|$ is the determinant of $\mathbf{x}$ and $\Gamma_k$ is the multivariate gamma function:
$$
\Gamma_{k}\left(\frac{n}{2}\right)=\pi^{k(k-1) / 4} \prod_{j=1}^{k} \Gamma\left(\frac{n}{2}-\frac{j-1}{2}\right).$$

Using the Wishart distribution as a prior can then be done in the same fashion as the gamma distribution.

\section{Interpretability of Precision Learning via Beta Residuals}\label{sec:beta_residual}

This section introduces the second novel contribution of this paper: using the \emph{beta residual} for fault-detection.

The previous section discussed performing FTC using precision learning. While this technique comes with the benefit of fault recovery without the residual thresholds, fault detection was never explicitly performed. Indications that a fault has occurred or is occurring may be useful for the user of the fault-tolerant controller. Common alternative fault-tolerant schemes explicitly detect faults to trigger recovery, providing the user with the controller's belief of the presence of faults. While precision learning does not yield outputs for explicit fault detection, the precision (inverse covariance) of the sensor model is estimated online at every time-step. This section introduces ways to combine explicit fault detection and isolation with precision learning, including a novel approach called the `beta residual`.

\subsection{Precision learning in conjunction with other methods}

A straightforward way to achieve fault detection explicitly is to use existing FDI methods in conjunction with precision learning. For instance, we can learn a probabilistically robust threshold from data offline (when the system is not operating) similar to approaches in \cite{baioumy2021ECC}. Then online (when the system is in operation) we compute the residual signal as  $\|y-\hat{x} \|$, where $y$ is the measurement of the sensor and $\hat{x}$ is the estimate of the state $x$. In our Bayesian Controllers it holds that: $\|y-\hat{x} \| = \|y-\mu_x \|.$

Now fault detection is straightforward to perform: the residual $\|y-\hat{x} \|$ is compared against the learned threshold and a fault is detected once the threshold is exceeded. Fault recovery can happen by triggering precision learning. 

This gives us two ways to achieve fault-tolerant control. Precision learning with \textit{implicit fault detection} means that we do not monitor the system and perform precision learning during the whole operation. This means there is no explicit fault detection. Additionally, precision learning with \textit{explicit fault detection} refers to monitoring the systems and only triggering precision learning once a fault is detected. The two methods are compared in the results section and the mean squared error is given in Table \ref{table:MSE_ft_control}.   

%\subsection{Stochastic Fault detection}

\subsection{The beta residual}
\label{subsec: beta residual}

As discussed, precision learning can work in conjunction with existing fault detection schemes. The controller uses the root mean sqaure error (RMSE) of a Kalman filter $\|y-\hat{x} \|$ as a residual signal and compares it to a learned threshold. We will now consider a different residual based on Bayesian inference, the \emph{beta residual}. 
 
The previous section discusses precision learning as Bayesian inference where instead of computing a point-estimate of the precision, we compute a full distribution. An appropriate choice in this case would be the Gamma distribution, as this is the conjugate prior precision to the Gaussian likelihood noise model with known mean and unknown precision (The Wishart distribution models the precision for the same reason in multi-dimensional cases). This Gamma distribution is parameterised by $\beta$ and $\alpha$. Since these parameters encode information about the degree to which a sensor is faulty, one of these parameters may be used as a residual signal. Given that $\alpha$ quantifies the number of observations, it does not correlate with the fault-status of a sensor on its own. Information about the sensor faultiness  will be rather encoded in $\beta$. The parameter $\beta$ is inversely proportional to the expected precision: 

\begin{equation}
    \mathbb{E} [\Gamma (\alpha, \beta)] = \frac{\alpha}{\beta}
\end{equation}

Now we will discuss how to extract interpretable outputs from the Gamma precision parameter $\beta$. Using sensor data labelled with the presence of faults, supervised learning algorithms can be applied to map from $\beta$ to a probability of a fault. For the sake of demonstration, we use a simple linear classification algorithm, logistic regression \cite{walker1967estimation}. 

Logistic regression assumes a linear relationship between the log-odds $\ln \frac{p}{1-p}$ and the predictor $x$: $\ln \frac{p}{1-p} = b x + b_0$, where $b$ and $b_0$ are model parameters selected to minimize the mean squared error of the classifier. By simple algebraic manipulation, $p = \frac{1}{1+e^{-(b x + b_0)}}$. The sigmoid function, $f(x) = \frac{1}{1+e^{-x}}$ constrains the output of the classifier to always be between 0 and 1. These output prediction are often interpreted as probabilities \cite{walker1967estimation}. A logistic regression model can easily be converted to a binary classifier by introducing a threshold on the model's output probability, such as all outputs $p > 0.6$ predicting a fault.

In summary, although precision learning does not explicitly determine the probability of a fault, simulated faults can generate the training data necessary for developing a classifier that uses learned precision statistics to return the probability of a fault. Combined with thresholds on the probability of a fault, the classifier can notify the user of a fault. While precision learning can be triggered using alternative FDI techniques, fault-tolerant control can also be achieved by always using precision learning and extracting interpretable sensor fault detections using beta residuals. This comes with the benefit of having fault classifications align with the controller's belief of the faultiness of sensors: When the sensor is believed to be operating correctly, the inferred precision is high (low variance), while when the sensor is believed to have a fault, the inferred precision will be low (high variance).
\section{Results}\label{sec:results}

% \textcolor{red}{We need to add 1) Some control performance stats/plots. This includes stuff like settling time, overshoot, RMSE from current state and goal ...etc. 2) Include clear details of how many sensors are present in each experiment, 3) Have at least one plot of the performance}

This section summarizes the results of experimentation in this paper. First, the fault tolerance of the u-AIC with point-mass precision learning is demonstrated. Then, the Bayesian Controller with precision learning as a distribution is shown to outperform alternative model-based FT techniques. Finally, a technique for extracting interpretable outputs from the Bayesian Controller classifying a fault is explained and applied.

\subsection{u-AIC with point-mass precision learning}
In the first experiment, we apply the methods in Sec.~\ref{subsec:unbiased_active_inference_controller} on a 2-DOF robotic manipulator. The manipulator has 3 sensor: a position encoder, a velocity encoder and a visual sensor. We use point-based precision learning on the u-AIC. 

We test two scenarios: a) precision learning at all times thus performing no explicit fault detection, b) precision learning only when a fault is detected. The first case has no explicit fault detection. In the second case, we use an existing FDI method based on the probabilistically-robust thresholds, then when a fault is detected, precision learning is triggered. 

In the simulations, the sensors are injected with zero-mean Gaussian noise. The standard deviation of the noise for encoders is set to $\sigma_q = \sigma_{\dot{q}}  =0.001$, while the one for the camera is set to $\sigma_{v} = 0.01$. The camera is also affected by barrel distortion with coefficients $K_1 = -1.5e^{-3},\ K_2=5e^{-6},$ $K_3=0$ (values are similar to work from \cite{cameraNoise,lipkin}). The agent starts in configuration $\bm x_0$, then moves to the targets $\bm x_1$ and $\bm x_2$. At $t = 8s$ a fault is injected. The encoder fault is such that the output related to the first joint freezes.

Without precision updates, the system is not able to reach the target state after the occurrence of the fault. Instead, the robot arm reaches a different configuration to minimise the free-energy, which is built fusing the sensor information from the (faulty) encoders and the (healthy) camera. However, when the faulty encoder is adjusted using precision learning, the agent is able to reach the final configuration. The Mean Squared Error (MSE) between the belief and the true position ($\bm \mu_x - \bm x$) is computed on a sample of test runs and reported in the Table~\ref{table:MSE_ft_control}. The results are reported for both the joint whose encoder is faulty, and joints with healthy encoders.

\begin{table}[ht]
\centering
\begin{tabular}{c|cc}
 &
  \begin{tabular}[c]{@{}c@{}}Joints with\\ encoder fault\end{tabular} &
  \begin{tabular}[c]{@{}c@{}}Joints without\\ encoder fault\end{tabular} \\ \hline
No fault-tolerance   & 0.0036    & 0.0020 \\
PL with implicit fault detection       & 5.422 e-5 & 4.527 e-5 \\
\begin{tabular}[c]{@{}c@{}}PL with explicit fault detection  \end{tabular} &
  6.097 e-5 &  4.134 e-5
\end{tabular}
\caption{Mean Squared Error (MSE) for different methods of fault-tolerant control. PL indicates precision learning}
\label{table:MSE_ft_control}
\end{table}

% The u-AIC with point-mass precision learning can perform stochastic FTC through sensor redundancy. This is a feedback controller that computes the most likely value for the state and the control action rather than a full posterior distribution. Additionally, to perform explicit fault detection, the RMSE of the Kalman filter is used as a residual $\|y-\hat{x} \|$. This controller is flexible, computationally fast and can be used on arbitrary systems with limited parameters that require re-tuning. 

\subsection{The Bayesian controller with precision learning}

In the second experiment, we compare precision learning on the Bayesian Controller to existing methods. Extensive simulations are performed on a cruise control to showcase the performance. The agent has two identical sensors, one of which suffers form a fault. We consider 3 types of fault: freezing, sensory injection and sensor drift. The agent is tasked with tracking 3 types of trajectories: constant, linear ramp and sinusoidal.

\subsubsection{Benchmarks}
To evaluate the performance of precision learning, two alternative techniques are applied as benchmarks: (Easy and Fast FDI) EF-FDI \cite{EasyAndFast5744127} and (State estimation residual FT) SER FT \cite{kommuri2016robust}. 

In EF-FDI, the detection procedure takes advantage of temporal redundancy in sensor observations, and is chosen here as it is representative of methods that are of a low complexity and fast computationally.

\begin{equation}
    r_k = |y_t - 2y_{t-1} + y_{t-2}|, \hspace{1mm}
    R_k = r_k + r_{k-1} + r_{k-2}
\end{equation}

\noindent where $y_t$ is the observation. If the statistic $R_k$ exceeds a threshold $\delta_{\text{EF}}$, the sensor output is ignored for the expected duration of the fault. 

Similarly, SER FT \cite{kommuri2016robust}, determines faults using analytical and sensor redundancy. A threshold $ \delta{\text{SER}}$ on the residual for state estimation $|y_t - \hat{x}_t|$ is computed. If this threshold is exceeded, the sensor is deemed faulty and the output is ignored. 

\subsection{Cruise Control System}

The Bayesian controller is tasked with tracking a trajectory for a vehicle cruise control problem given by the following system equations:

\begin{equation}
\bm x_{t+1} = (1 - \frac{b}{m} dt ) \bm x_t + \frac{dt}{m} \bm u_t + \bm q
\end{equation}

where $\bm x_t$ is the velocity of the car at time $t$, $b$ is the drag coefficient, $m$ is the mass of the car, and $\bm u$ is the control action and $\bm q$ is the process noise. Finally, $dt$ is the timestep size. The discrete observation model when no sensor faults are present is given by

\begin{equation}
\bm y_t = \bm x_t + \bm \eta
\end{equation}

where $\bm \eta$ is the observation noise. The exact parameter values used are provided in Table~\ref{table:cruise_control_parameters}.

\begin{table}[ht]
    \centering
    \begin{tabular}{ c|c|c } 
        Symbol & Name & Value \\
        \hline 
         $b$ & drag coefficient& $5 Ns/m$ \\
         $m$ & car mass & 100 $kg$ \\
         $dt$ & timestep size & 0.02 $s$ \\
         $\bm q$ & process noise & $\mathcal{N}(0, 0.001 m^2s^{-2})$ \\
         $\bm \eta$ & observation noise & $\mathcal{N}(0, 1 m^2s^{-2})$ \\
         \\ 
        
    \end{tabular}
\caption{System parameters for the vehicle cruise control problem.}
\label{table:cruise_control_parameters}
\end{table}

\subsubsection{Tracking Performance}

Applied to the same control problem, each FT technique is tested on 9 different ``scenarios'' for each FT technique (EF-FDI, No FT, Precision Learning (PL), and SER FT). The scenarios are described below.
% These 9 scenarios involve tracking 3 different trajectories: sinusoid, a ramp (linear trajectory) and a step response, and three different fault types: 
\begin{figure}[ht]
\centering
  \centering
  \includegraphics[width= 0.75 \linewidth]{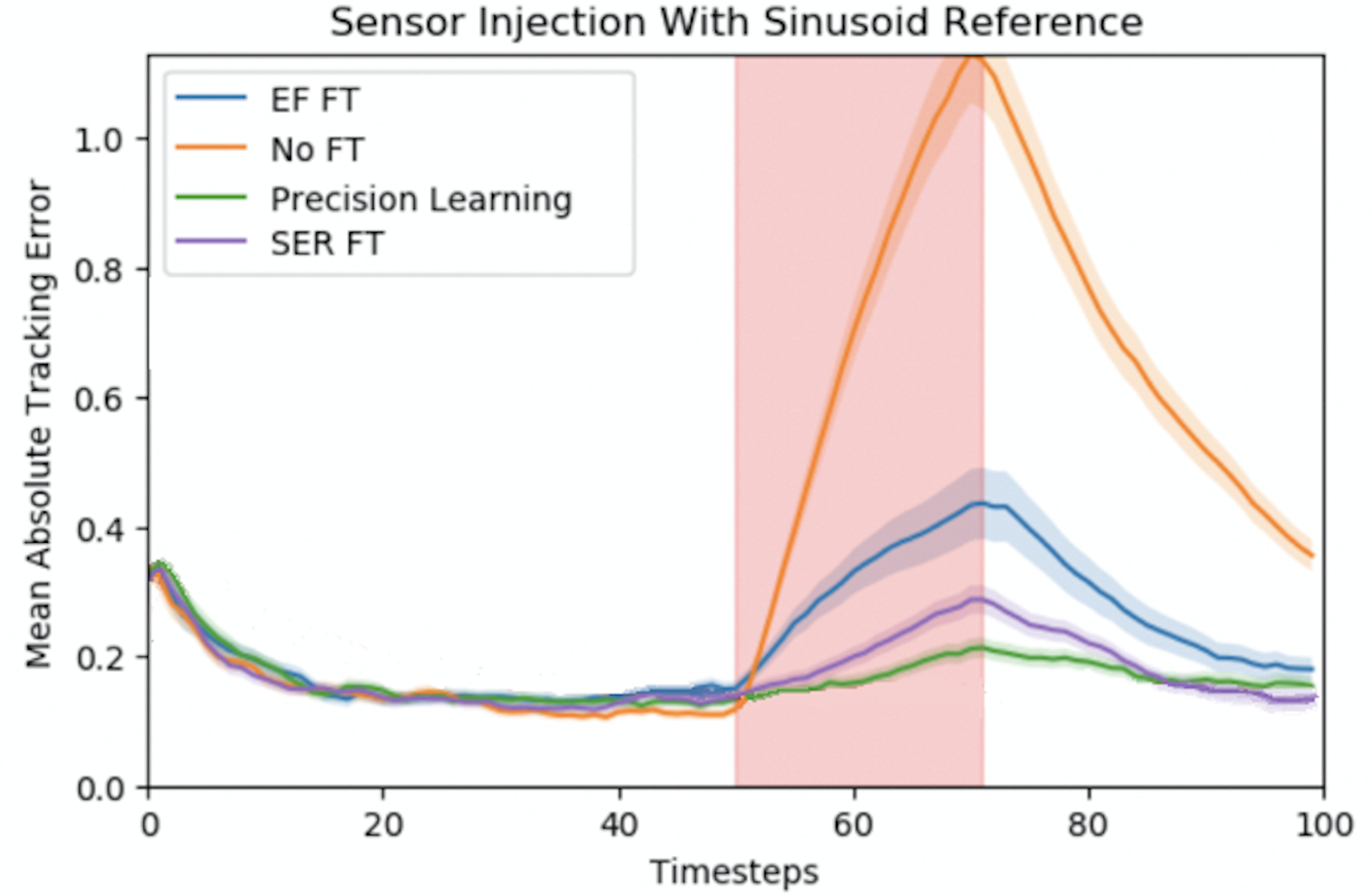}
  \caption{Mean absolute error in state, where reference is a sinusoid, and the sensor injection is present during the red highlighted period. }
  \label{subfig:sensor_injection_sinusoid_reference}
\end{figure}

To create a quantifiable summary of the relative merit of the different techniques across a range of scenarios, the following procedure was followed. The Bayesian controller is tasked with a cruise control problem across 900 simulations for each FT technique (EF-FDI, No FT, Precision Learning (PL), and SER FT). The simulations are distributed evenly across sensor fault type (sensor freeze, sensor drift, and sensor injection) and target state trajectory (constant, linear ramp, sinusoidal). Each unique combination of sensor fault type and target state trajectory will be referred to as a `scenario`. The mean root mean tracking error across 100 experiments for each scenario is compared using heteroskedastic $t$-tests for difference of means, summarized in Table~\ref{table:pre_post_fault_t_test}. Each element in the table shows the number of $t$-tests where the top FT technique's out-performance of the left technique was statistically significant minus the number of $t$-tests where the reverse was true. Tests were conducted across all three sensor failure types and all three reference paths, for a maximum of 9 in each cell (except for the ``Total Merit``). The total Merit entries are the sum of the column, showing total outperformances minus underperformances for a given technique, for a maximum total of 27 (9 per alternative technique). The ensemble mean absolute error across a single scenario, the sensor injection with sinusoid reference is shown in Figure \ref{subfig:sensor_injection_sinusoid_reference}.

% These results show that precision learning can be used effectively for fault-tolerant control. A Bayesian controller with precision learning is fast enough compared to the benchamarks, and consistently shows improved performance across different fault types and trajectories to follow. It also require no pre-processing (e.g. learning thresholds or tuning the controller to the type of trajectory to be encountered). 

\begin{table}[ht]
    \centering
    \begin{tabular}{ |c|c|c|c|c| } 
        \hline
         FT Technique & EF-FDI & No FT & PL & SER FT \\ \hline 
        EF-FDI & 0 & -6 & 8 & 8 \\ 
        No FT & 6 & 0 & 8 & 8  \\
        PL & -8 & -8 & 0 & -5  \\
        SER FT & -8 & -8 & 5 & 0  \\ \hline
        
        Total Merit & -10 & -22 & 21 & 11 \\ \hline 
    \end{tabular}
\caption{Comparison for different FTC methods with the relative merit reported. }
\label{table:pre_post_fault_t_test}
\end{table}

\subsection{Interpretability using Bayesian residual}

% When using a fault-tolerant controller, the user may want to be aware of the presence of faults. Typical FT techniques have separate FDI and fault recovery procedures in which FDI technique provides a clear indication that a fault is occurring. With precision learning, however, FDI and recovery are handled jointly through inference: there is no interpretable probability or classification of a fault. Intuitively, the inferred sensor precision should decrease when a sensor output is deemed less reliable. This subsection leverages the change in inferred sensor precisions to classify the faultiness of sensors using logistic regression on simulated data, and shows that with this extension, precision learning can be as interpretable as other FDI techniques.

In the third experiment, we compare the novel `beta residual' to the more common State Estimation Residual (equivalent to the RMSE of the Kalman filter).   

% We now consider the usage of a Bayesian residual for explicit fault detection. For comparison, the State Estimation Residual as an alternative predictor variable will be used. The statistic generated from Easy FT is not considered because the statistic is used to detect the \textit{onset} of a fault, rather than the \textit{presence} of a fault itself. The resulting logistic regression model returns an interpretable probability of the presence of a fault, and can be converted into a classifier by setting a threshold on the output probability.

In the procedure, 1000 sensor faults were generated, with a mean-time-to-fault of 2.8 s, and mean-time-to-recovery of 0.4 s for a total of approximately 160,000 simulation timesteps. The occurrences of the faults and recoveries are distributed assuming a constant probability of failure. The faults are therefore Poisson-distributed in their time-to-occurence and in length. When a fault is triggered in the simulation, a random selection of one of the 3 faults is selected. The model assumes that only one fault at a time occurs with the same sensor. Faults are allowed to occur simultaneously at both of the sensors (although this is not common). The learned precision parameters and the SER residual are saved together with the binary fault truth values. The results are split into 80\% training set and 20\% test set. Logistic regression is trained on the training set and evaluated on the test set. 

\begin{figure}[ht]
\centering
  \centering
  \includegraphics[width= 0.8 \linewidth]{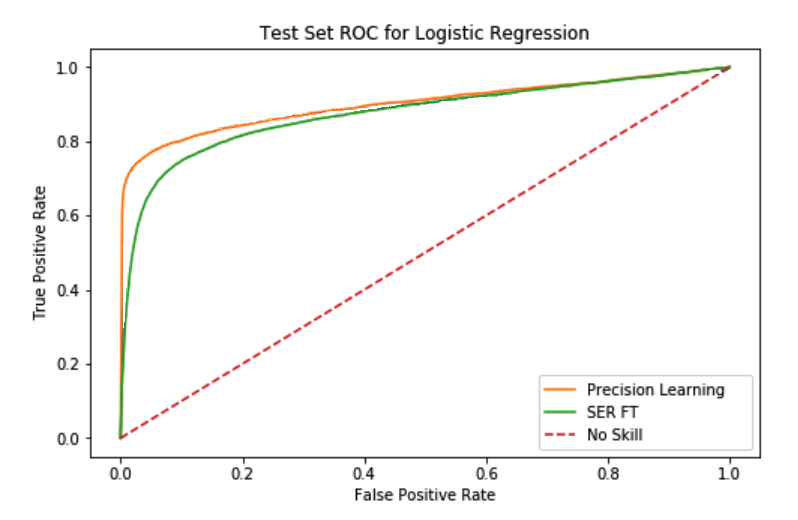}
  \caption{The Receiver Operating Characteristic (ROC) shows the trade-off between true positive rate and false positive rate for all possible thresholds. The diagonal line in red is the ROC of a hypothetical model that guesses randomly. EF-FFT is not shown for comparison because its residual for fault detection corresponds with the onset of faults, not the presence of faults, rendering it unsuitable for classification using logistic regression. }
  \label{subfig:lookback_vs_expanding_computation}
\end{figure}

The test set Receiver Operating Characteristic (ROC) is shown in Table \ref{fig:lookback_vs_expanding_data}. The ROC-AUC is the are under the ROC. It is a common measure of the performance of binary classifiers that return probabilities rather than classes. For the controller with the given FT strategy on the left, the model input is used to predict the 0-1 faultiness of the sensor with logistic regression. The ROC-AUC, an assessment of the classification power of the model.

A brief explanation of the ROC as a metric follows. In order to evaluate a classifier, accuracy is an insufficient metric, due to imbalances in the classes (there are more non-faulty cases than faulty cases). In the case of binary classification, both the true positive rate and the false positive rate should be considered. However, in order to evaluate the logistic regression model as a classifier, a probability threshold for classification must be selected. The ROC evaluates the model at all possible thresholds: Each point on the curve represents a threshold with an associated false positive rate ($x$-axis) and true positive rate ($y$-axis). The area under the curve measures the fit of the logistic regression model: The closer to 1, the better the classifier. A random guesser has an ROC-AUC (Area Under Curve) of 0.5, labelled as ``No Skill`` in Figure \ref{subfig:lookback_vs_expanding_computation}.

These results show that the parameter beta can be effectively used as a residual signal. A stochastic threshold was learned using logistic regression, however, other supervised models may be used.

\begin{table}[ht]
    \centering
    \begin{tabular}{ |c|c|c| } 
        \hline
        FT Technique & Input & AUC \\
        \hline 
        SER FT & $\|y-\hat{x} \|$ & 0.873 \\ 
        Precision Learning & $\beta$ & 0.907 \\ 
        \hline
    \end{tabular}
\caption{ROC-AUC for fault detection with different residuals.}
\label{fig:lookback_vs_expanding_data}
\end{table}
\section{Discussion and conclusion}\label{sec:conclusions}

This paper compares a set of fault-tolerant controllers based on Bayesian inference and on a novel approach: precision learning. All controllers perform state-estimation, control, fault detection and recovery as a single inference procedure. Sensor faults are modelled as changes in the covariance/precision of the sensor model. Thus by learning the precision online, the agent can achieve fault-tolerant control. In addition, by modeling the control problem as Bayesian inference, standard methods using non-Gaussian distributions and non-linear systems can be leveraged. The results show how controllers based on precision learning outperform existing approaches on a variety of tasks. This includes different types of fault (sensor freeze, sensor injection and drift), while tracking different trajectories (ramp, step-response, and sinusoid). 

A key limitation of this approach is the lack of an explicit fault detection step. To mitigate this issue, one could use standard fault detection approaches in conjunction with precision learning. We demonstrate that this yield satisfactory performance. Additionally, we introduce a novel \emph{beta residual}, which shows improved performance compared the commonly used residuals such as the root mean square error (RMSE) of a Kalman filter. Additionally, it is pointed out how the beta residual can be interpreted as an hyper-parameter which represents the trust in a sensor at a specific time instant. Future work will further investigate the use of the beta residual for fault-detection, including its fault detectability properties.

% We show that using this beta residual allows for improving fault detection compared to commonly used root mean square error (RMSE) of the Kalman filter

%\bibliographystyle{IEEEtran}
\bibliography{IEEEabrv,bibFile}

\end{document}